\documentclass[preprint]{revtex4-2}

\usepackage{graphicx}
\usepackage{dcolumn}
\usepackage{bm}
\usepackage{hyperref}
\hypersetup{
    colorlinks=true,       
    linkcolor=blue,        
    citecolor=blue,        
    filecolor=magenta,     
    urlcolor=blue          
}
\usepackage[version=4]{mhchem}
\usepackage{amsmath,amssymb,amsfonts}%
\usepackage{siunitx}%
\DeclareSIUnit{\wv}{\%\,w/v}
\DeclareSIUnit{\square}{sq}
\DeclareSIUnit{\pix}{pixel}
\DeclareSIUnit{\fps}{fps}
\DeclareSIUnit{\radians}{rad}
\DeclareSIUnit{\elecVolt}{eV}

\begin{document}

\preprint{}

\title{Density modulations in active colloidal systems through orthogonal propulsion control and sensory delays} 

\author{Ueli T\"opfer}


\author{Maximilian R. Bailey}

\author{Sanjay Schreiber}
\author{Federico Paratore}
\thanks{Corresponding author}
\email{federico.paratore@mat.ethz.ch}
\author{Lucio Isa}
\thanks{Corresponding author}
\email{lucio.isa@mat.ethz.ch}

\affiliation{%
Department of Materials, ETH Zurich, Zurich, Switzerland
}%

\date{\today}

\begin{abstract}
Recent advancements in active colloidal systems aim to mimic key characteristics of biological microswimmers, particularly their adaptive motility in response to environmental changes. While many approaches rely on externally imposing a propulsive force, achieving true autonomous and self-regulating adaptation to the environment remains limited. In this study, we develop and analyze Janus microswimmers driven by electrohydrodynamic flows that autonomously adjust their propulsion dynamics in response to varying illumination conditions. Our Janus particles are silica colloids partially coated with titania, which self-propel via induced-charge electrophoresis (ICEP) under uniform AC electric fields. Since titania is photoconductive, it increases its conductivity under UV illumination, which thereby regulates the propulsion velocity independently of and orthogonally to the applied electric field. Crucially, the velocity adaptation requires a finite time. This sensory delay, which we systematically characterize, leads to enhanced microswimmer localization in response to spatiotemporal light modulations compared to the typical case of instantaneous response considered for synthetic microswimmers. By harnessing these dynamics, akin to those of biological microswimmers, we exert precise control over both local and global particle behavior, presenting novel opportunities for adaptive active matter systems.

\end{abstract}

\maketitle

\section{Introduction} \label{sec:intro}

Spontaneous spatial and temporal variations of population density in response to environmental inputs are central to life at all length scales, from survival strategies in populations of complex organisms, to embryonic development and biofilm formation in cellular colonies. The ability to sense, elaborate and actively respond to stimuli and cues, from visual perception to light or chemical fields, is an essential feature of adaptive behavior. Individual units regulate their functions and thus contribute to coordinated collective responses.
At the microscale, some motile organisms are capable of adapting their motility in response to chemical fields \cite{Bhattacharjee2021} in order to accumulate in regions of high nutrient concentration \cite{Wadhams2004}, to proliferate in areas with defined oxygen levels \cite{Lefvre2014, Frankel2009}  or to actively seek natural light \cite{Garcia2013,DiLeonardo2024} 

In synthetic microscale active systems, such as ensembles of self-propelling Janus particles, local density modulations can be used to template materials \cite{Palacci2013}, assemble small-scale devices \cite{Aubret2021, Ma2015}, regulate flows \cite{Bricard2013} and carry out complex manipulation tasks \cite{Lffler2023, Heuthe2024}. While physical interactions can lead to structure formation \cite{Yan2016} and density variations, i.e. including motility-induced phase separation \cite{Cates2015,Geyer2019}, density variations can also spontaneously emerge as a consequence of spatial motility modulations even in the absence of specific interactions among the active units. 

A general theoretical result in fact determines that $\rho(r) \propto \frac{1}{v(r)}$, where $\rho(r)$ is the steady-state mean number density of microswimmers at position $r$ and $v(r)$ is the corresponding local propulsion speed \cite{Schnitzer1993,Stenhammar2016,Tailleur2008}. The result is exact for non-interacting objects with a position-dependent swimming speed that is independent of local density, with corrections to account for those effects. Nonetheless, this description makes the crucial assumption that the velocity changes instantaneously as a function of position. 
While this condition may be implemented in numerical simulations \cite{Stenhammar2016}, any real system will have a finite response time.

This velocity adaptation can then be described by introducing a response time $\tau$, defined as the characteristic time to update propulsion speed in response to sensing an input. In most experimental realizations of artificial microswimmers with propulsion control $\tau$ is assumed to be negligible.
This assumption holds if two conditions are verified: i) $\tau \ll D_\text{R}^{-1}$, where $D_\text{R}^{-1}$ represents the characteristic Brownian reorientation time, implying that the direction of motion does not change before the velocity $v$ is updated, and ii) $\tau\,v \ll R$ implying that the swimmer travels a small distance compared to its radius $R$ during the time $\tau$, which is the typical case \cite{Lavergne2019, Zehavi2022, GomezSolano2017}. However, there are few exceptions in which delayed responses are deliberately engineered to alter steady-state swimmer density distributions \cite{FernandezRodriguez2020}. 

In contrast, biological microswimmers often exhibit significantly longer response times due to the involvement of complex biochemical signaling pathways that lead to velocity adaptations. As a result, the distance traveled before the velocity changes, $\tau v$, frequently exceeds the swimmer size ($\tau v > R$), and the response time may be either shorter or longer than the reorientation timescale, $\tau \gtrless \tau^\text{reorient}$ \cite{Arlt2018, Frangipane2018}. An intermediate and highly tunable case is the one of macroscopic robots, where response parameters can be fully programmed. In such systems, the distance traveled during velocity adaptation typically exceeds the size of the robot ($\tau\,v > R$), while the relation to the rotational timescale can be freely adjusted to either suppress or enhance density modulations, depending on the desired behavior \cite{Mijalkov2016, Leyman2018}.

In principle, density variations can derive from any mechanism providing local propulsion control. Nonetheless, in practice, light-controlled microswimmers have emerged as an optimal solution due to the ease of spatial and temporal modulation of light intensity patterns via digital projection devices and spatial light modulators. Light typically acts as the stimulus responsible for the propulsion, e.g. by driving ionic fluxes across the cell wall of motile bacteria 
\cite{Walter2007} or due to photo-thermal effects, either directly via temperature gradients \cite{Jiang2010,Qian2013} or by coupling those to chemical \cite{Lozano2016} or surface tension gradients \cite{Dietrich2020}. Therefore, in these cases, light is the sole control parameter in the system. Having a single control parameter limits both the versatility and the robustness of applicable strategies to regulate motility, and correspondingly density modulations, within independently available environmental energy sources. An exception may be considered the one of photo-catalytic microswimmers \cite{Singh2018,Wang2018, Bailey2021}, where the particles actively decompose environmentally available chemical fuel at an illumination-dependent rate. However, independently and dynamically controlling fuel concentration levels is challenging.

A preferred alternative would instead require the realization of systems with propulsion control mechanisms orthogonal to the mechanism generating it, very much like a sailboat that can regulate speed independently of the wind by changing the surface of its sails. Whereas examples exist in biological swimmers, for instance, positive chemotactic response toward non-metabolizable attractants, even against nutrient gradients \cite{Zhang2019, Sourjik2001}, or gravitaxis observed under constant nutrient conditions \cite{Giometto2015}, strategies for orthogonal propulsion control in synthetic systems are far less frequent.
In particular, shape reconfiguration \cite{Ohiri2018,Dou2019,Archer2020} and external regulation of material properties, \cite{Zehavi2022}, or both \cite{vanKesteren2023}, have been proposed in the literature as single-particle-level solutions, opening up exciting opportunities to implement new strategies for global control of particle density.  

In this work, we report a strategy to regulate the spatial distribution of photo-responsive silica-titania Janus microparticles powered by uniform AC electric fields via the orthogonal control of the conductivity of the titania caps through UV illumination. We discuss the resolution of particle localization, or patterning, as a function of the sensory delay imposed by the properties of the titania cap. From these results we draw guidelines to dynamically engineer the localization of synthetic microswimmers by the selection of materials with tailored responses.

\section{Results and Discussion} \label{sec:results}
\subsection{Propulsion adaptation}
\begin{figure}[h]%
\centering
\includegraphics[width=0.9\textwidth]{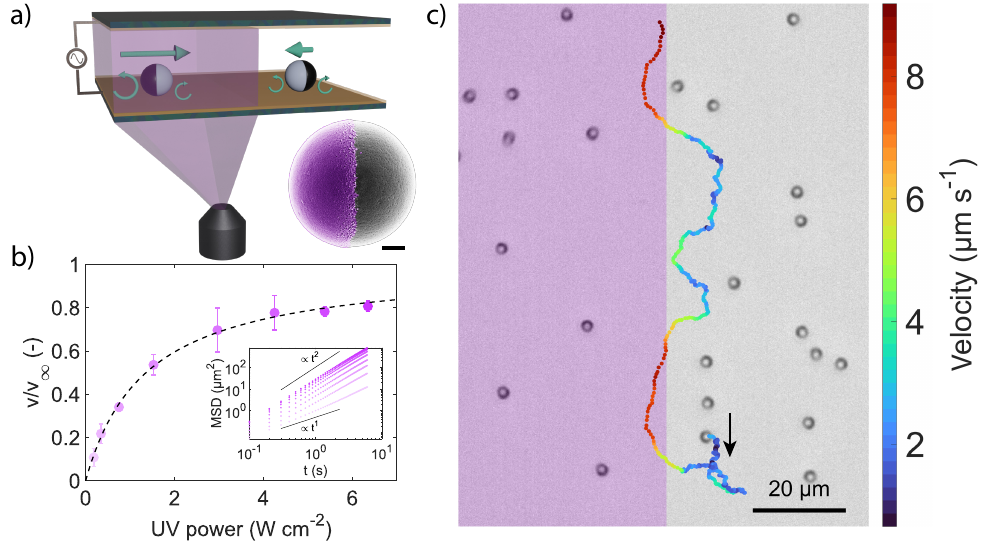}
\caption{\textbf{Adapting active particle velocity via light modulation.} a) Schematic of an experimental cell illustrating the setup for our experiments. The curled green arrows represent the electro-hydrodynamic flows around the Janus particles, and which are modulated by the UV light pattern projected through the microscope objective via a digital micromirror device (DMD). Inset: scanning electron micrograph of an individual Janus particle with the titania cap false-colored in violet. The scale bar corresponds to \SI{500}{\nano\meter}. b) Particle velocity $v$ (\SI{5}{\volt} and \SI{4}{\kilo\hertz}) normalized by its saturation value $v_{\infty}$ as a function of UV light intensity. Inset: ensemble-averaged MSD curves for light intensity levels corresponding to the colors of the data points in the main graph. c) Optical micrograph and particle trajectory (\SI{10}{\volt} and \SI{10}{\kilo\hertz}) in a binary on-off stripe illumination pattern (illuminated area colored in violet; \SI{6.35}{\watt\per\centi\meter\squared}). The color bar represents the particle velocity, calculated from the displacements over \SI{300}{\milli\second}.
}\label{fig:1}
\end{figure}

Our particles are fabricated from silica colloids of \SI{2.06}{\micro\meter} diameter that are half-coated with a \SI{50}{\nano\meter}-thick layer of titania (Figure \ref{fig:1}a). These Janus particles exhibit self-propulsion through induced-charge electrophoresis (ICEP) \cite{MBoymelgreen2022,squires_bazant_2006}, which is based on the contrast in conductivity between the coated and bare hemisphere \cite{Boymelgreen2012}. The titania, in its anatase phase, exhibits photoconductivity, implying that its electrical conductivity increases upon illumination with photons of wavelengths corresponding to the energy of the bandgap or higher ($\lambda \leq \SI{387}{\nano\meter}$) \cite{Luttrell2014, BRAJSA2004151}, thus enabling control over the conductivity contrast of the Janus particles via UV illumination \cite{Zehavi2022}. As shown in Figure \ref{fig:1}a, the particles are confined in a liquid layer formed between two parallel and transparent electrodes, and, due to gravity, settle onto the bottom one. Upon applying a alternating (AC) electric field in the kHz frequency range, ICEP induces an asymmetric flow field around each particle that propels them in a direction parallel to the electrode surface. As expected for ICEP, we measure swimming velocities proportional to the square of the applied electric field, given as $v \propto \left(\frac{V_\text{pp}}{h}\right)^2$ \cite{SQUIRES2004}, as shown in Supplementary Fig. 1. Here, $V_\text{pp}$ represents the amplitude of the applied voltage (varied experimentally between \SIrange{1}{10}{\volt}) and $h$ is the electrode spacing (fixed to \SI{120}{\micro\meter} in our setup). 

We control particle velocity $v$ by modulating the illumination intensity within the cell using light with a wavelength of \SI{365}{\nano\meter} and a power density ranging from \SIrange{0}{6.35}{\watt\per\centi\meter\squared}. As displayed in Figure \ref{fig:1}b, for fixed electric fields (\SI{5}{\volt} at \SI{4}{\kilo\hertz}), the normalized velocity first increases linearly with increasing intensity before asymptotically approaching a plateau value ($v_{\infty}$) at higher intensities. This follows a Michaelis-Menten-like kinetic, as previously reported for the propulsion velocity of photocatalytic microswimmers \cite{Palacci2013,Chen2016, Aubret2018}. We extract these active velocities from ensemble-averaged mean squared displacements (MSD) of Janus particles subjected to globally uniform illumination of varying intensities (Figure \ref{fig:1}b inset), following the approach outlined in subsection \ref{subsec:experiments}.

Using a digital micromirror device (DMD) placed within the optical path of the microscope, we then project light patterns across the field of view, thus creating tailored spatiotemporal modulations of light intensity in which the particles move. Figure \ref{fig:1}c presents a typical trajectory of a particle navigating a binary stripe illumination pattern, alternating between \SI{0}{\watt\per\centi\meter\squared} (gray) and \SI{6.35}{\watt\per\centi\meter\squared} (violet) regions. The particle's velocity varies significantly, from less than \SI{1}{\micro\meter\per\second} in dark areas to greater than \SI{8}{\micro\meter\per\second} in illuminated regions. Correspondingly, its persistence length, $L_\text{p} = v / D_\text{R}$, which is the average distance a particle moves before de-correlating its orientation due to its rotational diffusivity $D_\text{R}$, is substantially higher upon UV illumination. However, from the qualitative color change in the trajectory, one observes that the velocity adaptation is not instantaneous upon crossing the illumination boundary. 

\subsection{Sensory delay}
\begin{figure}[h]%
\centering
\includegraphics[width=0.99\textwidth]{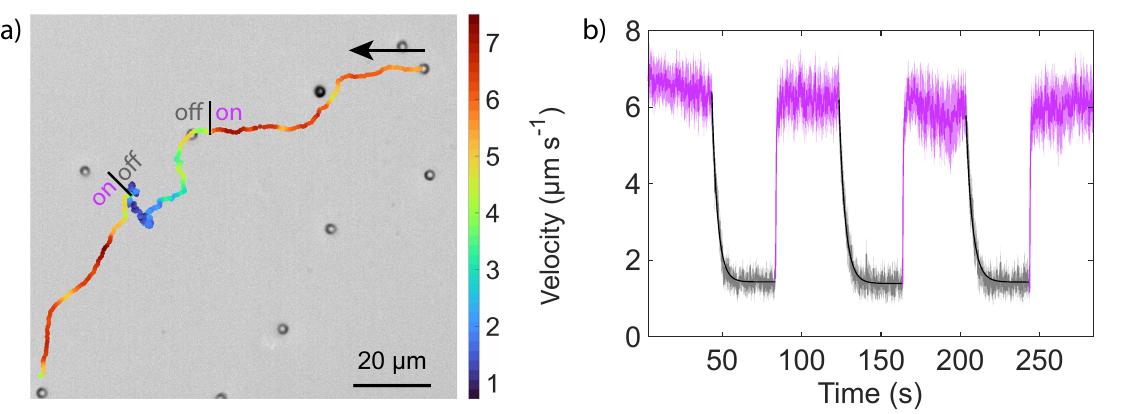}
\caption{\textbf{Finite sensory delay} a) Optical micrograph showing the trajectory of a single particle undergoing a cyclic switch of the global illumination from \SIrange{6.3}{0}{\watt\per\centi\meter\squared} and back to \SI{6.3}{\watt\per\centi\meter\squared} (with $V_{pp}=$\SI{5}{\volt} at \SI{4}{\kilo\hertz}). The color bar represents the particle velocity, calculated from the displacement over \SI{300}{\milli\second}. b) Ensemble-averaged swimming velocity of 144 particles subjected to multiple cyclic changes of the global illumination (with $V_{pp}=\, \SI{5}{\volt}$ at \SI{4}{\kilo\hertz}). The black lines represent exponential fits to the velocity transitions between the ON and OFF illumination states (eq. \ref{eq:exp_fit}). The velocities were calculated from the displacement over \SI{300}{\milli\second} and averaged across the particle ensemble, with the shaded area representing the standard deviation. }\label{fig:2}
\end{figure}

To investigate the time dependence of the velocity adaptation to changes in illumination, we therefore subjected the particles to cycles of ON and OFF illumination, i.e. switching the  illumination intensity between \SI{6.3}{\watt\per\centi\meter\squared} and \SI{0}{\watt\per\centi\meter\squared}. Figure \ref{fig:2}a shows a typical particle trajectory during such a cycle, where we observe the persistence of higher propulsion speed after the UV light is turned off and a rapid acceleration upon turning it on again. We then quantified this response by measuring the  ensemble-averaged velocity for all particles in the sample. Figure \ref{fig:2}b reveals a strong asymmetry in the response times: the particles exhibit a finite delay when transitioning from high to low activity while they immediately adapt from low to high activity.

We quantified these adaptation timescales by fitting the change in velocity with an exponential decay function:

\begin{equation}
\label{eq:exp_fit}
    v(t) = v_{i} + \left(v_{i-1}-v_{i}\right)e^{\frac{-t}{\tau}}
\end{equation}

In this equation, $ v(t) $ is the velocity at a given time $ t $, $ v_{i-1} $ and $ v_i $ are the velocities at the end of the previous illuminated and current dark phase, calculated as the average velocity over \SI{20}{\second}, respectively. $ \tau $ is the characteristic response time which is the sole fitting parameter. This quantification reveals that the response time to slow down is on the order of \SI{2}{\second}, while the response to increasing illumination is essentially instantaneous on the experimental timescale ($\tau \lesssim$ exposure time). Delays on the order of seconds are usually not explicitly considered in photoactive colloidal systems and are often treated as instantaneous \cite{Dong2017}. However, as we will see later, they play a crucial role in defining the steady-state particle distribution under position-dependent illumination. 
 
While charge carrier lifetimes on the order of seconds are unusually long for semiconducting materials, studies have reported photo-generated electron lifetimes ranging from seconds to minutes. These extended lifetimes are attributed to the stabilization of surface-trapped holes through hydrogen-bonding interactions between adsorbed water and \ce{TiO2} \cite{Litke2017a, Litke2017b}. 

This response is very different from the reaction of the particles to switching on and off the AC electric field while the illumination intensity is kept constant, where the adaptation is symmetric and instantaneous (Supplementary Fig. 2).

\subsection{Localization}

\begin{figure}[h]%
\centering
\includegraphics[width=0.99\textwidth]{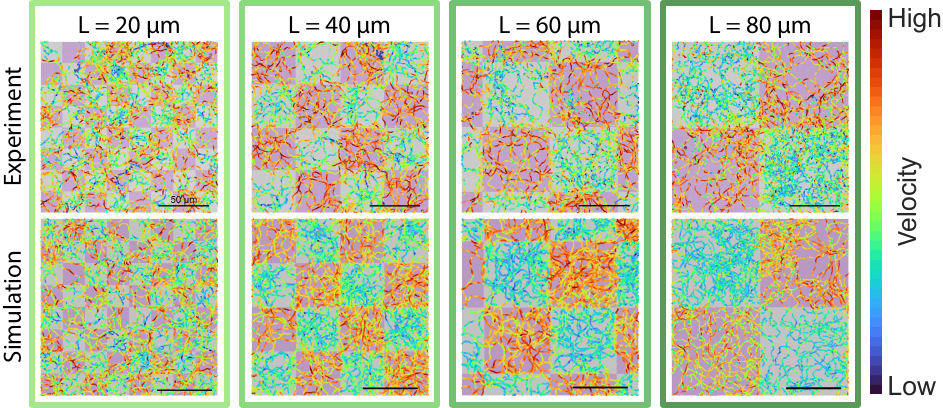}
\caption{\textbf{Particle motion on checkerboard motility patterns.} Top row: Experimental particle trajectories within illumination checkerboard patterns (\SI{0}{\watt\per\centi\meter\squared}--gray and \SI{6.3}{\watt\per\centi\meter\squared}--violet) with varying size L at \SI{4}{\volt} and \SI{4}{\kilo\hertz}. Bottom row: Simulated particle trajectories with input parameters extracted from the corresponding experiments. Instantaneous velocities are color-coded between $v_L$ and $v_H$.
}\label{fig:3}
\end{figure}

We investigate the combined effect of spatial and temporal modulation of particle motility via light signals as a mechanism to control the global density distribution of particles \cite{Caprini2022, Sker2021}. Using a DMD, we project checkerboard patterns of ON (\SI{6.3}{\watt\per\centi\meter\squared} -- violet) and OFF (gray) regions of characteristic size $L$, corresponding to regions of high and low particles' activity, respectively. The  top row of Figure \ref{fig:3} shows experimental trajectories, color-coded by velocity, for particles moving through patterns with increasing $L$ from \SIrange{20}{80}{\micro\meter}.

In order to ensure that we have a conceptual understanding of the crucial factors at play, we replicate the experimental conditions using a numerical simulation framework previously developed for spatial modulations of rotational dynamics \cite{FernandezRodriguez2020} and active velocity \cite{Bailey2024}. In brief, the simulation solves the Langevin equations of motion for non-interacting particles with position-dependent velocity. Specifically, each particle has a defined minimum (low) and maximum velocity (high), $v_L$ and $v_H$, respectively, and switches between those two at a rate proportional to $\frac{1}{\tau}$, where $\tau$ is the characteristic response time (more details in Section \ref{subsec:numsim}). By selecting values of $v_L$, $v_H$ and $\tau$ extracted from the experiments, the corresponding simulated trajectories (color coded following the experimental procedure) show good qualitative agreement with the experimental ones (Figure \ref{fig:3} bottom).

\begin{figure}[h]%
\centering
\includegraphics[width=0.55\textwidth]{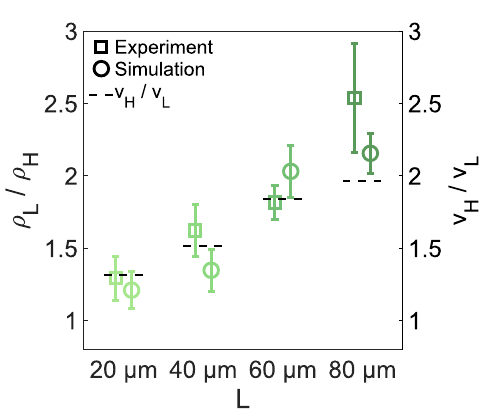}
\caption{\textbf{Quantification of particle localization} Comparison of the localization, defined as the ratio of particle density in the low- and high-activity areas $\rho_L / \rho_H$, between the experiments (squares) and simulations (circles) for different pattern sizes and velocity ratios. The dashed lines correspond to $\rho_L v_L = \rho_H v_H$.
}\label{fig:4}
\end{figure}

We quantify the particle number density in non-illuminated $\rho_L$ areas and illuminated $\rho_H$ ones by counting the particles in the respective regions in every frame. As initially described, at steady state, we expect to see a localization of particles in lower-velocity regions according the theoretical results of \cite{Schnitzer1993,Cates2015,Stenhammar2016}. We quantify the degree of localization as the ratio between $\rho_L$ and $\rho_H$, with $\rho_L / \rho_H = 1$ corresponding to no localization.
For systems with velocity adaptation we thus expect localization to follow the simple relation $\rho_L v_L = \rho_H v_H$, where $v_L$ and $v_H$ are the two corresponding swimming speeds. In our experiments and simulations we indeed recover this proportionality for small pattern sizes (Figure \ref{fig:4}). However, upon increasing the velocity ratio $v_H / v_L$ and the pattern size $L$, we achieve an "over-localization" of particles in the low-motility regions, effectively allowing for a more efficient accumulation of the particles within the slow areas. In order to further elucidate how this behavior depends on the parameters of the systems, and in particular on the response time $\tau$, we then carried out a set of systematic numerical simulations. 

\begin{figure}[h]%
\centering
\includegraphics[width=0.95\textwidth]{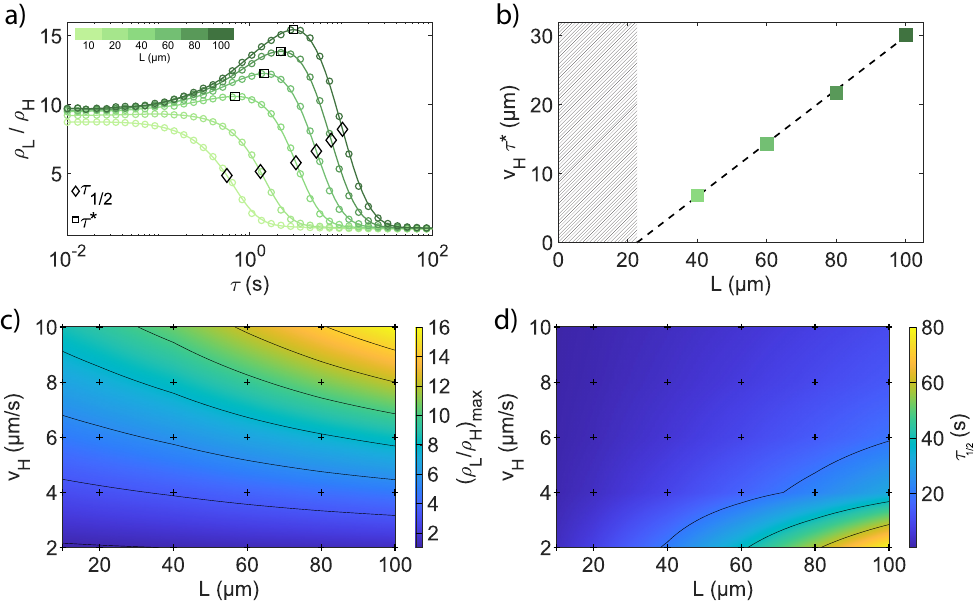}
\caption{\textbf{Contrast and resolution in particle patterns from localization.} a) Localization $\rho_L / \rho_H$ as a function of sensory delay $\tau$ for checkerboard patterns with sizes ranging from $L = \SIrange{10}{100}{\micro\meter}$. $v_L$ and $v_H$ are set to $\SI{1}{\micro\meter\per\second}$ and $\SI{10}{\micro\meter\per\second}$, respectively. Continuous lines represent smoothing splines fitted to individual simulations (colored circles). We extract two characteristic parameters from each simulation: $(\rho_L / \rho_H)_\text{max}$, the maximum localization obtained at the optimal sensory delay $\tau^*$  for each $L$ (squares), and $\tau_\text{1/2}$, the delay time $\tau$ at which localization efficiency drops below $\SI{50}{\percent}$ of the maximum (diamonds). b) Optimal adaptation length scale, $v_H \tau^*$, as a function of pattern size  $L$. The dashed line represents a linear fit to data points for $L = \SIrange{40}{100}{\micro\meter}$ and the hatched area are shows where the optimal delay drops below 0 ($\tau^*<0$). c) Map of $(\rho_L / \rho_H)_\text{max}$ for $L = \SIrange{10}{100}{\micro\meter}$ and $v_H = \SIrange{2}{10}{\micro\meter\per\second}$. The continuous map is interpolated between experimental data points (black dots). d) Map of $\tau_\text{1/2}$ for the same parameters as in c.}\label{fig:5}
\end{figure}

Figure \ref{fig:5}a shows the density ratio as a function of sensory delay for a fixed minimum ($v_L$) and maximum velocity ($v_H$) of $\SI{1}{\micro\meter\per\second}$ and $\SI{10}{\micro\meter\per\second}$ for different values of $L$. In particular, for $L$ = \SIrange{20}{40}{\micro\meter} we start observing a non-monotonic relation between the response time $\tau$ and the localization ratio  $\rho_L / \rho_H$. While for fast enough response timescales $\tau \lesssim \SI{0.1}{\second}$ the expected localization relation $\rho_L v_L = \rho_H v_H$ is recovered for all values of $L$, localization is significantly enhanced for L$\geq$\SI{40}{\micro\meter} at an optimal delay time for which $\rho_L / \rho_H$ reaches a maximum. This maximum as well as the dependence on $L$ can be qualitatively understood using simple scaling arguments. Ballistic particles, when transitioning from $v_H$ to $v_L$ regions, can penetrate into the low motility region by a length corresponding to $\min\left(v_H\tau, v_H / D_\text{R}\right)$ that is limited by the smallest of two time scales: the delay time $\tau$ and the Brownian rotational time $1/D_\text{R}$ (for the investigated experimental system $\tau < 1/D_\text{R}$ ). The deeper the particles move ballistically into the low motility region the longer it takes them to leave with the low velocity $v_L$. This sensory delay thus causes the emergence of an optimal delay time, $\tau^*$, which increases with increasing pattern size $L$. If the delay is further increased the localization starts to drop as the particles cannot adapt to velocity changes on a length scale $L$ and velocity modulations become decoupled from the underlying motility landscape \cite{FernandezRodriguez2020}. 

We analyze the dependence of the optimal delay time on the pattern size by comparing the traveled distance before adaptation at $\tau^*$, $v_H \tau^*$, to the pattern size $L$ (Figure \ref{fig:5}b). For large enough patterns $L \geq \SI{40}{\micro\meter}$, this distance increases linearly with the pattern size, as indicated by the linear regression. 
This dependence can be quantified by computing the average length $L^\text{eff}$ of a ballistic trajectory crossing a square domain of size $L$. The calculation requires integrating over all possible trajectories that enter on one side of the square and exit through one of the other three (see Supplementary section S3). This integration gives $L^\text{eff} \approx 0.82 L$. Correspondingly, localization is maximized for a penetration inside the low-motility regions of $0.5\,L^\text{eff} \approx 0.41\,L$, close to the fitted slope of 0.39 in Figure \ref{fig:5}b. Thus, this optimal penetration distance follows $v_H \tau^* = 0.5 \left(L^\text{eff}\,- \,L^\text{eff}_{\tau^*=0}\right)\, \approx 0.41\left(L\,- \,L_{\tau^*=0}\right)$, where $L_{\tau^*=0}$ denotes the smallest pattern size for which a finite sensory delay enhances localization (i.e., when $\tau^* > 0$). This threshold size corresponds to the persistence length of the particle in the slow region, $v_L \tau_\text{R}$, which represents the ballistic escape distance beyond which delayed adaptation no longer provides a benefit. In our case $L^\text{eff}_{\tau^*=0} = 2\,v_L\,\tau_R$ and thus $ L_{\tau^*=0} \approx \SI{14}{\micro\meter}$ in reasonable agreement with the fitted value of \SI{23}{\micro\meter}.

Previous work on photokinetic bacteria, showed that illumination patterns can be used to control density distributions creating bacterial "images" via motility control \cite{Frangipane2018,Arlt2018}. In the spirit of providing quantitative guidelines for pattern optimization for our synthetic microswimmers, we identify two parameters for every given combination of minimum and maximum velocity, and length scale, which effectively correspond to a maximum contrast in the pattern and to a delay time above which only \SI{50}{\percent} of the maximum contrast is achieved for a given pixel size $L$. To this end, we simulated a variety of systems with $v_H = \SIrange{2}{10}{\micro\meter\per\second}$, $L = \SIrange{10}{100}{\micro\meter}$ and a fixed $v_L = \SI{1}{\micro\meter\per\second}$. The first parameter is in fact the maximum achievable localization $(\rho_L / \rho_H)_\text{max}$. From the data in Figure \ref{fig:5}a, we observe that maximum contrast is achieved for instantaneous adaptation at small pixel sizes but takes place at finite values of $\tau$ for larger $L$. Figure \ref{fig:5}c thus shows how much can the contrast be enhanced at a given $L$ by choosing the optimum sensory delay $\tau^*$, demonstrating the existence of an interplay between maximum contrast and pixel size as a function on the particle response time. The second parameter is $\tau_\text{1/2}$, which is the sensory delay above which localization drops below half of its maximum (Figure \ref{fig:5}a). 
As we show in Figure \ref{fig:5}d, $\tau_\text{1/2}$ remains below \SI{20}{\second} for a broad range of values and only only significantly increases for a combination of low $v_H$ and large $L$ as only a combination of the two increases the residence time of the particles in a given area of the pattern Figure \ref{fig:5}d. These two parameters can thus be applied to govern efficient localization in an active colloidal system as they define an achievable contrast ($\left(\rho_L / \rho_H\right)_\text{max}$) and pixel size ($L$) at which this contrast can be obtained, given a minimum and maximum velocity and as a function of the sensory delay. Their identification thus guides the search for materials with prescribed value of $\tau$ or clarifies which contrast and spatial resolution can be achieved for a given system.

\section{Conclusions} \label{sec:discussion}

Our results demonstrate that adaptive motility in synthetic active colloids can be achieved by incorporating materials that intrinsically modulate their properties in response to environmental stimuli, decoupled from the driving mechanism for propulsion. In our case, while both the AC electric field and the illumination are externally applied, the propulsion adaptation is an intrinsic response of the microswimmers, dictated by the choice of materials.

The use of light to modulate motility, combined with a digital micromirror device (DMD), allows precise, independent control of numerous particles at the microscale, enabling mimicking behaviors of biological microswimmers, such as taxis \cite{berg1975chemotaxis, adler1975chemotaxis,Autrum1979, Jkely2008, Marcos2012}. Moreover, real-time tracking and the use feedback loops based on particle positions open the door to develop complex control strategies. On the one hand, those can be used to obtain dynamically self-organized systems to perform coordinated tasks \cite{Vutukuri2020, Wang2020, Lffler2023, Lavergne2019} and to optimize navigation strategies \cite{Cichos2020, Vizsnyiczai2017, Pellicciotta2023, Putzke2023}. Conversely, one can implement feedback strategies inspired by natural behavior to introduce effective signaling interactions, where the motility of certain particles in the system is influenced by the behavior of others. Among those cases, visual perception models can be implemented \cite{Lavergne2019} or, in the future, local particle accumulation or depletion may be triggered upon reaching a given density, similar to the quorum sensing behavior of certain microorganisms \cite{MorenoGmez2023}.

Looking forward, the integration of dynamic responses in the development of control strategies for microscale active systems will present exciting opportunities. The presence of finite, characteristic adaptation times has already been used to enhance contrast in microswimmer patterns \cite{Frangipane2018}, to rectify bacterial motion \cite{MassanaCid2022},  and to control robotic swarms \cite{Leyman2018, Mijalkov2016}. The central role of sensory delay, as demonstrated in this work, may guide the selection of materials with specific time responses in order to target specific individual or collective behavior. In electrically-powered swimmers, this can be achieved by tailoring the charge-carrier lifetime, for example by stabilizing holes on the surface \cite{Litke2017b} or by introducing a hole scavengers such as an alcohol \cite{Mohamed2011}. 
The choice may be extended to other cues beyond light, such as temperature, chemical concentration or pH \cite{Heinen2017} to perform complex tasks in diverse environments \cite{Lanz2024,Yoshida2020} in an increasingly more autonomous fashion.
Finally, the presence of finite delays for particles to slow down upon removing UV light, presents analogies with underactuated control systems, where dynamics may continue after cutting off an energy input. This behavior is suggestive of analogies with what is known as passive dynamics in robotics, which constitutes an important family of control strategies to optimize navigation and energy consumption \cite{Collins2005}.

In conclusion, we envisage the rapid evolution of adaptive microscale systems where the convergence of materials' selection and discovery with control systems will pave the way to the development of  colloidal materials with precise microscale control and localization without external computerized intervention.
\section{Methods} \label{sec:methods}

\subsection{Particle Fabrication}
For the fabrication of the adaptive silica-titania Janus particles, silica colloids (\SI{2.06}{\micro\meter} diameter, \SI{5}{\wv}, microParticles GmbH, Germany) are first diluted to \SI{0.5}{\wv} in MilliQ water. Following this \SI{100}{\micro\litre} of the sonicated and diluted colloidal suspension are spread on a glass slide that was previously cleaned using a \SI{1}{\min} air-plasma treatment. The suspension is then let to dry to form a colloidal monolayer. The monolayer is then coated with \SI{50}{\nano\meter} of \ce{Ti3O5} using an e-beam evaporator (Plassys MEB700SL) and annealed at \SI{500}{\degreeCelsius} for \SI{}{min} in a rapid thermal processor (AS-One 150, ANNEALSYS). Finally, the particles are retrieved from the slides by sonication in MilliQ water. The formation of a majority of anatase-phase \ce{TiO2} was confirmed through X-ray powder diffraction (supplementary Fig. 4). Prior to an experiment, the particle dispersion is mixed in equal amounts with a \SI{0.2}{\wv} Pluronic F-127 solution to render a particle dispersion with \SI{0.1}{\wv} of the surfactant. This surfactant 
reduces the adhesion of the particles to the substrate of the cell. 

\subsection{Cell preparation}
The transparent conductive electrodes (\SI{16}{\milli\meter} x \SI{16}{\milli\meter}) consist of indium-tin oxide coated glass ($<\SI{7}{\ohm\per\square}$, pr\"azisions glas \& optik). Two of these electrodes are sandwiched together with a \SI{120}{\micro\meter} thick adhesive spacer with a \SI{9}{\milli\meter} circular opening (Grace Bio-Labs SecureSeal, USA). The spacer was halved to provide an in- and outlet through which \SI{9.5}{\micro\litre} of the particle suspension is filled into the cavity before being closed off with grease (Krytox GPL205). The two electrodes are then contacted with adhesive copper tape and connected to a signal generator (33522A Arbitrary Waveform Generator, Agilent) that applies the AC electric field with varying voltages between \SIrange{1}{10}{\volt} at frequency between \SIrange{4}{10}{\kilo\hertz}.

\subsection{Experiments}
\label{subsec:experiments}
All experiments are conducted on an inverted optical microscope (Eclipse Ti2-E, Nikon) with magnifications of 40x (CFI S Plan Fluor ELWD 40XC, Nikon, Japan) and 60x (CFI S Plan Fluor ELWD 60XC, Nikon, Japan). The intensity of the UV light ($\lambda =$ \SI{365}{\nano\meter}, UHP-F-365, Prizmatix) can be varied from \SIrange{0}{6.35}{\watt\per\centi\meter\squared}. The illumination is patterned within the microscope's field of view using a digital micromirror device pattern illuminator (Polygon 1000, Mightex, USA). A DAPI filter cube is used to remove most of reflected UV light and avoid oversaturation of the camera. Image sequences (\qtyproduct{2160 x 2560}{\pix}) of various lengths are obtained with a sCMOS camera (Andor Zyla) at \SI{10}{\fps}. The particle positions in each frame are detected and linked to trajectories using various custom-build Matlab routines.

The translational diffusion coefficient $D_\text{T}$ for each experimental cell was obtained by fitting the ensemble-averaged MSD in the absence of an applied electric field with the theoretical MSD for a particle undergoing Brownian motion:

\begin{equation}
\label{equ:msd_diff}
    \text{MSD}(\tau) =4D_\text{T}\tau
\end{equation}

Active particle velocities were obtained by fitting ensemble-averaged MSD curves with the theoretical MSD model for active particles given by \cite{Howse2007, Bechinger2016}:

\begin{equation}
    \label{equ:msd_active}
    \text{MSD}(\Delta t) = \left(4D_\text{T} + 2\,v^2D_\text{R}^{-1}\right)\tau + \frac{2\,v^2}{D_\text{R}^{2}}\left(e^\frac{-\Delta t}{D_\text{R}^{-1}}-1\right)
\end{equation}

In this equation, $D_\text{T}$ and $D_\text{R}$ represent the translational and rotational diffusion coefficients, respectively, $v$ is the active particle velocity, and $\Delta t$ is the elapsed time difference time. The value of $D_\text{T} \approx \SI{0.1}{\micro\meter\squared\per\second}$ for each cell was determined from the Brownian motion observed in the absence of an electric field, as described above. The rotational diffusion coefficient, $D_\text{R}$, was set to the theoretical value for spherical particles in water at room temperature ($D_\text{R}^{th} = \frac{k_\text{B}T}{8\pi\eta R^3} = \SI{0.160}{\radians\squared\per\second}$), consistent with the values obtained from mean squared angular displacement curves for various illumination intensities and electric field strengths (supplementary Fig. 5). 

\subsection{Numerical simulations}
\label{subsec:numsim}
The numerical model used for the simulations was previously developed in \cite{Bailey2024} and is based on solving the following equations of motion:

\begin{equation}
\begin{aligned}
m \ddot{x} & =f_x(\mathbf{r}, \tau, \theta)-\gamma_T \dot{x}+\sqrt{2 k_B T \gamma_T} \eta_x(t) \\
m \ddot{y} & =f_y(\mathbf{r}, \tau, \theta)-\gamma_T \dot{y}+\sqrt{2 k_B T \gamma_T} \eta_y(t) \\
I \ddot{\theta} & =\gamma_R \dot{\theta}+\sqrt{2 k_B T \gamma_R} \eta_\theta(t)
\end{aligned}
\end{equation}

where the force acting on the particles is given by:

\begin{equation}
\begin{aligned}
f_x(\mathbf{r}, \tau, \theta) & =V(\mathbf{r}, \tau) \cos (\theta) \gamma_T \\
f_y(\mathbf{r}, \tau, \theta) & =V(\mathbf{r}, \tau) \sin (\theta) \gamma_T
\end{aligned}
\end{equation}

Here $m$ and $I$ are the mass and moment of inertia of the particles (density $\rho = \SI{2.5}{\gram\per\cubic\centi\meter}$, radius $r = \SI{1}{\micro\meter}$). $\gamma_T$ and $\gamma_R$ are the translational and rotational friction coefficients and $V(\mathbf{r}, \tau)$ is the response and position depend velocity. Finally $\eta_x\left(t\right)$, $\eta_y\left(t\right)$ and $\eta_\theta\left(t\right)$ are uncorrelated noise terms that fulfill: 

\begin{equation}
\begin{aligned}
& \left\langle\eta_x\right\rangle=\left\langle\eta_y\right\rangle=\left\langle\eta_\theta\right\rangle=0 \\
& \left\langle\eta_x^2\right\rangle=\left\langle\eta_y^2\right\rangle=\left\langle\eta_\theta^2\right\rangle=1
\end{aligned}
\end{equation}

The maximum velocities used as input to the simulations were sampled from a skew-normal distribution fitted to the velocity distribution observed in high-activity areas during the experiments. Corresponding minimum velocities were calculated from the maximum velocities using the ratio of active velocities between illuminated and non-illuminated areas, derived from MSD curves fitted with equation \ref{equ:msd_active}.

\section*{Acknowledgments}
L.I., U.T., F.P. and M.R.B. acknowledge funding from the European Research Council (ERC) under the European Union’s Horizon 2020 Research and Innovation Program grant agreement No 101001514. The authors also acknowledge Dr. Thomas Weber from the D-MATL X-Ray Service Platform at ETH Zurich for performing X-ray measurements, as well as the use of cleanroom facilities at the Binnig and Rohrer Nanotechnology Center (BRNC) at IBM Research – Zurich and at the FIRST Center for Micro- and Nanoscience at ETH Zurich. The authors thank C. van Baalen for helpful comments and support during the review of the article.

\section*{Author Contributions}
Author contributions are defined based on the CRediT (Contributor Roles Taxonomy) and listed alphabetically. 
Conceptualisation: U.T., L.I., F.P.; 
Data Curation: U.T.; 
Formal Analysis: U.T.; 
Funding Aquisition: L.I.; 
Investigation: U.T., F.P., S.S; 
Methodology: U.T., F.P., L.I.; 
Project Administration: L.I.; 
Resources: U.T.; 
Software: U.T., M.R.B.; 
Supervision: L.I., F.P.; 
Validation: U.T.; 
Visualisation: U.T., F.P., L.I.; 
Writing - original draft: U.T.; 
Writing - review and editing: U.T., F.P., L.I.;

\section*{Conflicts of interest}
There are no conflicts to declare.

\bibliography{references.bib}

\newpage

\section*{S1: Particle velocity as a function of the electric field}
\renewcommand{\thefigure}{S1}
\begin{figure}[h]
\centering
\includegraphics[width=0.5\textwidth]{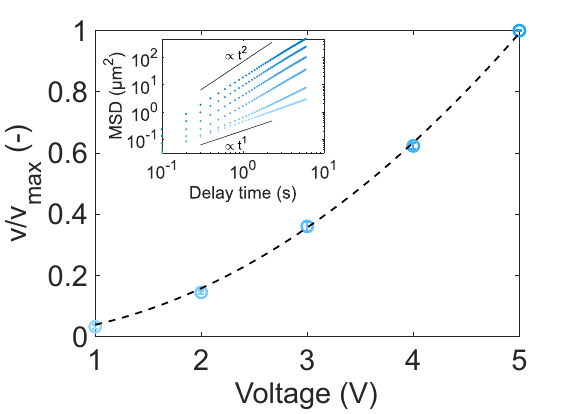}
\caption{\textbf{Normalized particle velocity as a function of the applied potential} Particle velocity increases quadratically with increasing applied voltage. The values were averaged over three measurements obtained at constant illumination intensity of \SI{6.3}{\watt\per\centi\meter\squared} and AC frequency of \SI{4}{\kilo\hertz} and normalized by the maximum value. The active velocities were calculated from ensemble averaged MSD curves as shown in the inset obtained at a constant illumination intensity of \SI{6.3}{\watt\per\centi\meter\squared}.}
\end{figure}

\newpage

\section*{S2: Velocity adaptation under switched electric field}
\renewcommand{\thefigure}{S2}
\begin{figure}[h]
\centering
\includegraphics[width=0.5\textwidth]{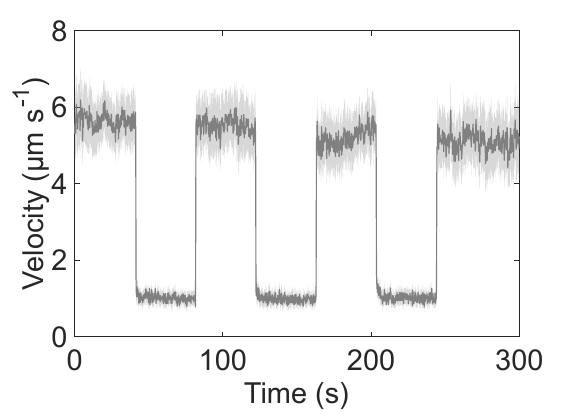}
\caption{\textbf{Velocity adaption under switched electric field} Ensemble-averaged swimming velocity of particles subjected to multiple cyclic changes of the electric field (\SI{5}{\volt}/\SI{0}{\volt}, \SI{4}{\kilo\hertz}) under constant illumination. The velocities were calculated from the displacement over \SI{300}{\milli\second} and averaged across the particle ensemble, with the shaded area representing the standard deviation. The illumination intensity was kept constant at \SI{6.3}{\watt\per\centi\meter\squared}.}
\end{figure}

\newpage

\section*{S3: Average trajectory length}

The average trajectory length, $L^\text{eff}$, of a particle traversing a square of size $L$ can be determined by integrating over all possible paths where the particle enters through one side and exits through any of the other three sides. This calculation can be decomposed into contributions from trajectories that exit through the opposite side and those that exit through one of the adjacent sides. The latter contribution is accounted for twice, corresponding to exits through either of the two adjacent sides:

\begin{equation}
\label{eq:traj_len}
\begin{aligned}
L^\text{eff} &= \frac{1}{3} \left(t^\text{\,top} + 2\,t^\text{\,side} \right) \\
&=  \frac{1}{3} \int_0^L \int_0^L\left(\sqrt{(x-y)^2+L}+2 \sqrt{x^2+y^2}\right) \,dx\,dy \approx 0.82\,L 
\end{aligned}
\end{equation}

\renewcommand{\thefigure}{S3}
\begin{figure}[h]
\centering
\includegraphics[width=0.5\textwidth]{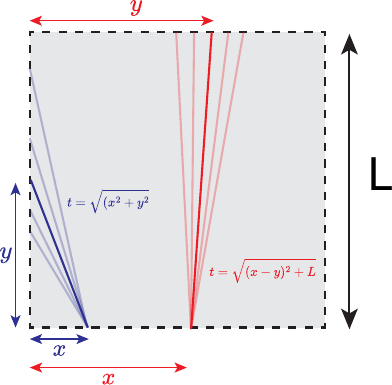}
\caption{\textbf{Average trajectory length through a square} Sketch to visualize the calculation of the length of a trajecotry entering the square and leaving it through the side (blue) or the top (red).}
\end{figure}

\newpage

\section*{S4: XRD measurement of \ce{TiO2}}

For the measurement, \ce{Ti3O5} was deposited on a silicon wafer following the same protocol as for the particle coating (see Methods).

\renewcommand{\thefigure}{S4}
\begin{figure}[h]
\centering
\includegraphics[width=0.5\textwidth]{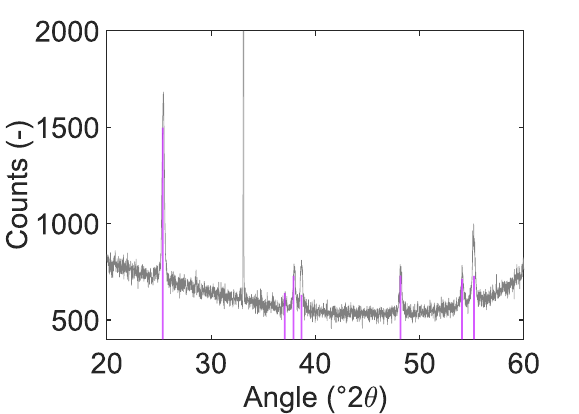}
\caption{\textbf{XRD spectrum of titania} X-ray powder diffraction spectrum of the annealed \ce{Ti3O5} and reference peaks of anatase titania (American Mineralogist Crystal Structure Database 0010735) shown in gray and violet, respectively.}
\end{figure}

\newpage

\section*{S5: Rotational diffusion under varied voltage and illumination}

Values for the rotational diffusion coefficient were extracted from mean squared angular displacement (MSAD) curves by fitting them with $\text{MSAD}(\tau) =2D_\text{R}\tau$, assuming Brownian dynamics with one rotational degree of freedom. The orientation of the particles was obtained by placing an axis connecting the center of the particle to the center of the cap. The center of the cap was determined in a second image analysis step with a more stringent thresholding.

\renewcommand{\thefigure}{S5}
\begin{figure}[h]
\centering
\includegraphics[width=0.99\textwidth]{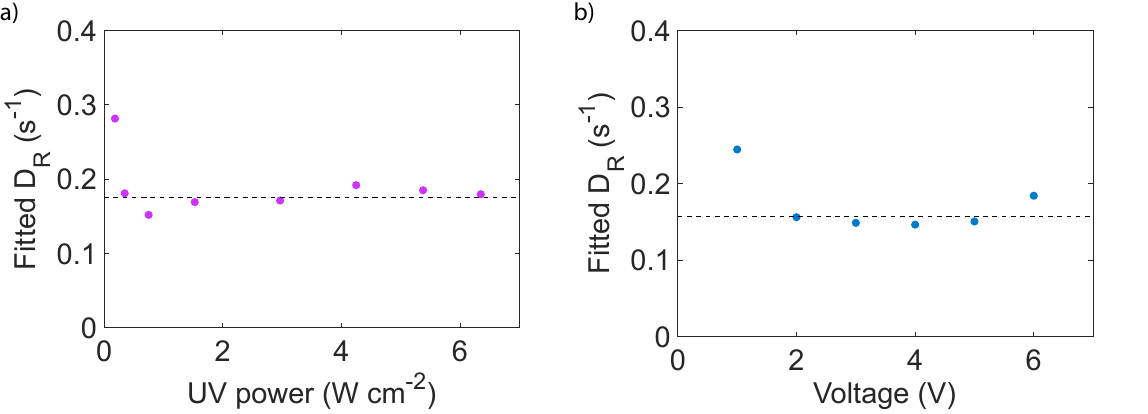}
\caption{\textbf{Influence of illumination and voltage on rotational dynamics} a) Rotational diffusion coefficient $D_\text{R}$ as a function of illumination intensity. The dashed gray line indicates the average value of \SI{0.1759}{\per\second}. The voltage was kept constant at \SI{4}{\volt} and \SI{4}{\kilo\hertz} throughout the experiment. The value at \SI{0.18}{\watt\per\centi\meter\squared} was excluded from the averaging as a minimum intensity was required to ensure a particle orientation perpendicular to the substrate and thus only one rotational degree of freedom. b) Rotational diffusion coefficient $D_\text{R}$ as a function of applied potential. The dashed line corresponds to a mean value of $D_\text{R} = \SI{0.1574}{\per\second}$. The illumination intensity was kept constant at \SI{6.45}{\watt\per\centi\meter\squared} throughout the experiment. The value for a potential of \SI{1}{\volt} was excluded with the same reasoning as in a).}
\end{figure}

\newpage

\section*{S6: List of Supplementary Videos}
\begin{itemize}
    \item \textbf{Video S1:} Time-lapse video of a particle ensemble subjected to multiple cycles of global illumination modulation between \SI{6.3}{\watt\per\centi\meter\squared} and \SI{0}{\watt\per\centi\meter\squared}, with an applied electric field of $V_{pp} = \SI{5}{\volt}$ at \SI{4}{\kilo\hertz}. The color bar indicates the particle velocity, calculated from the displacement over a \SI{300}{\milli\second} time window. The video corresponds to Figure 2 in the main text and is shown at 5× real-time speed. \\
    \item \textbf{Video S2:} Time-lapse video of a particle ensemble navigating a checkerboard illumination pattern, with alternating illuminated (violet overlay, \SI{6.3}{\watt\per\centi\meter\squared}) and non-illuminated (gray) regions. An electric field of $V_{\mathrm{pp}} = \SI{6}{\volt}$ at \SI{4}{\kilo\hertz} is applied throughout the experiment. The color bar indicates the particle velocity, calculated from the displacement over a \SI{300}{\milli\second} time window. The video is shown at 8× real-time speed.  \\
    \item \textbf{Video S3:} Time-lapse video of simulated particle trajectories with spatially varying velocities, based on input data extracted from the experiment shown in Video S2 (see Methods section for details). Regions of high particle velocity are shown with a violet overlay, while low-velocity regions appear in gray. The color bar indicates the instantaneous particle velocity, calculated from displacement over a \SI{300}{\milli\second} time window. The video is shown at 8× real-time speed.
\end{itemize}

\end{document}